# Bayesian and Hybrid Cramer-Rao Bounds for QAM Dynamical Phase Estimation

J. Yang, B. Geller, and A. Wei [1]

*Abstract*—In this paper, we study Bayesian and hybrid Cramer-Rao bounds for the dynamical phase estimation of QAM modulated signals. We present the analytical expressions for the various CRBs. This avoids the calculation of any matrix inversion and thus greatly reduces the computation complexity. Through simulations, we also illustrate the behaviors of the BCRB and of the HCRB with the signal-to-noise ratio.

## I. INTRODUCTION

There are three types of estimators widely used in communication systems [1]: data-aided (DA), code-aided (CA) and non-data-aided (NDA) estimators. DA estimation techniques obtain the better performance but may lead to unacceptable losses in power and spectral efficiency. CA synchronization allows an improved data efficiency but requires additional interactive impairments between decoding and synchronization units. Finally, NDA synchronization algorithms may sometimes lead to poor results but they exhibit the highest transmission efficiency and are still attractive.

To know whether an imposed algorithm is good or not, one often compares its performance with lower bounds. Although there exists many lower bounds, the Cramer-Rao bound (CRB) is the most commonly used and the easiest to determine [2], [3].

Many works [4]-[10] concern the CRBs for the carrier phase and frequency estimation in DA and NDA scenarios. But all these papers refer to an idealized situation in which the phase offset is constant. However, in modern burst-mode communications, a time-varying phase noise due to the oscillator instabilities has to be considered [11], [12]. Taking into account such a phase noise variance, [13] considered the DA CRB for the phase estimation with a phase noise variance and [14] derived a BCRB for NDA BPSK signals; [15] added the consideration of a deterministic parameter to obtain the HCRB for BPSK signals. However, the synchronization of BPSK signals is relatively simpler than that of QAM signals. The goal of this paper is then two-folded. First we detail the derivation of the BCRB and of the HCRB for QAM modulated signals. Second we want to provide benchmarks for the QAM dynamical phase estimation, since we present a generalization of the off-line synchronizing scheme [16] to QAM modulated signals in [17] whose performance can exactly reach the bounds derived in this paper. For lack of space reasons, this study is extended to the code-aided case in [18].

This paper is organized as follows. In section II, we recall the various kinds of Cramer-Rao bounds. After describing the system model in section III, we derive the BCRBs and the HCRBs for both on-line and off-line estimations in section IV. Moreover, we also present the analytical expressions for the various CRBs; this avoids computing the inverse of any information matrix. A discussion about the various CRBs and a conclusion are respectively provided in section V and VI. [2]

## II. CRAMER-RAO BOUNDS (CRBs) REVIEW

It is known that the parameters to be estimated can be categorized as deterministic or random parameters. Denote this parameter vector as $\mathbf{u} = (\mathbf{u}_r^T, \mathbf{u}_d^T)^T$, where $\mathbf{u}_d$ is a $(n-m) \times 1$ deterministic vector and $\mathbf{u}_r$ is a $m \times 1$ random vector with an a priori probability density function (pdf) $p(\mathbf{u}_r)$. The true value of $\mathbf{u}_d$ will be denoted $\mathbf{u}_d^\Delta$ and $\hat{\mathbf{u}}(\mathbf{y})$ is the estimator of $\mathbf{u}$ where $\mathbf{y}$ is the observation vector. The HCRB satisfies the following inequality [15] on the MSE:

$$E_{\mathbf{y},\mathbf{u}_r|\mathbf{u}_d=\mathbf{u}_d^\Delta}\left[(\hat{\mathbf{u}}(\mathbf{y})-\mathbf{u})(\hat{\mathbf{u}}(\mathbf{y})-\mathbf{u})^T_{|\mathbf{u}_d=\mathbf{u}_d^\Delta}\right] \geq \mathbf{H}^{-1}(\mathbf{u}_d^\Delta), \quad (1)$$

where $\mathbf{H}(\mathbf{u}_d^\Delta)$ is the so-called hybrid information matrix (HIM) and is defined as:

$$\mathbf{H}(\mathbf{u}_d^\Delta) = E_{\mathbf{y},\mathbf{u}_r|\mathbf{u}_d=\mathbf{u}_d^\Delta}\left[-\Delta_\mathbf{u}^\mathbf{u} \log p(\mathbf{y},\mathbf{u}_r|\mathbf{u}_d)_{|\mathbf{u}_d=\mathbf{u}_d^\Delta}\right]. \quad (2)$$

It is shown in [16] that inequality (1) still holds when the deterministic and the random parts of the parameter vector are dependent. By expanding $\log p(\mathbf{y},\mathbf{u}_r|\mathbf{u}_d)$ as $\log p(\mathbf{y}|\mathbf{u}_r,\mathbf{u}_d) + \log p(\mathbf{u}_r|\mathbf{u}_d)$, the HIM can be rewritten as:

$$\mathbf{H}(\mathbf{u}_d^\Delta) = E_{\mathbf{u}_r|\mathbf{u}_d=\mathbf{u}_d^\Delta}\left[\mathbf{F}(\mathbf{u}_d^\Delta,\mathbf{u}_r)\right] + E_{\mathbf{u}_r|\mathbf{u}_d=\mathbf{u}_d^\Delta}\left[-\Delta_\mathbf{u}^\mathbf{u} \log p(\mathbf{u}_r|\mathbf{u}_d)_{\mathbf{u}_d=\mathbf{u}_d^\Delta}\right], (3)$$

where $\mathbf{F}(\mathbf{u}_d^\Delta,\mathbf{u}_r) = E_{\mathbf{y}|\mathbf{u}_r,\mathbf{u}_d=\mathbf{u}_d^\Delta}\left[-\Delta_\mathbf{u}^\mathbf{u} \log p(\mathbf{y}|\mathbf{u}_d,\mathbf{u}_r)_{|\mathbf{u}_d=\mathbf{u}_d^\Delta}\right] \quad (4)$

is the Fisher information matrix (FIM).

In particular, if $\mathbf{u} = \mathbf{u}_d$, (4) reduces to:

$$\mathbf{H}(\mathbf{u}_d^\Delta) = \mathbf{F}(\mathbf{u}_d^\Delta) = E_{\mathbf{y}|\mathbf{u}_d=\mathbf{u}_d^\Delta}\left[-\Delta_{\mathbf{u}_d}^{\mathbf{u}_d} \log p(\mathbf{y}|\mathbf{u}_d)_{|\mathbf{u}_d=\mathbf{u}_d^\Delta}\right]. \quad (5)$$

The inverse of $\mathbf{H}(\mathbf{u}_d^\Delta)$ in (5) is just the standard CRB [2].

On the contrary, if $\mathbf{u} = \mathbf{u}_r$, (4) becomes:

---

[1] This work was partially funded by the ANR LURGA program.
J. Yang is with SATIE, ENS Cachan (email: yang@satie.ens-cachan.fr);
B. Geller is with LEI, ENSTA PARISTECH (email: geller@ensta.fr);
A. Wei is with LATTIS, Université Toulouse II (email: anne.wei@lattis.Univ-toulouse.fr).

[2] The notational convention adopted is as follows: italic indicates a scalar quantity, as in $a$; boldface indicates a vector quantity, as in $\mathbf{a}$ and capital boldface indicates a matrix quantity as in $\mathbf{A}$. The $(m,n)^{th}$ entry of matrix $\mathbf{A}$ is denoted as $[A]_{m,n}$. The transpose matrix of $\mathbf{A}$ is indicated by a superscript $\mathbf{A}^T$, and $|\mathbf{A}|$ is the determinant of $\mathbf{A}$. $\mathbf{a}_m^n$ represents the vector $[a_m,\cdots,a_n]^T$, where $m$ and $n$ are positive integers ($m < n$). $\text{Re}\{a\}$ and $\text{Im}\{a\}$ are respectively the real and imaginary parts of $a$. $E_{xy}[\ ]$ denotes the expectation over $x$ and $y$. $\nabla_\mathbf{u}$ and $\Delta_\mathbf{u}^\mathbf{v}$ represent the first and second order derivative operators.

$$\mathbf{H} = E_{\mathbf{u}_r}\left[\mathbf{F}(\mathbf{u}_r)\right] + E_{\mathbf{u}_r}\left[-\Delta_{\mathbf{u}_r}^{\mathbf{u}_r} \log p(\mathbf{u}_r)\right], \quad (6)$$

where $\mathbf{F}(\mathbf{u}_r) = E_{\mathbf{y}|\mathbf{u}_r}\left[-\Delta_{\mathbf{u}_r}^{\mathbf{u}_r} \log p(\mathbf{y}|\mathbf{u}_r)\right]$, (7)

and the inverse of $\mathbf{H}$ in (6) is the Bayesian CRB (BCRB) [3].

### III. SYSTEM MODEL

We consider the transmission of a sequence $\mathbf{s} = [s_1, \cdots, s_L]^T$ of M-QAM signals from a set $\mathbf{S}_M$, rotated by some random carrier phases $\boldsymbol{\theta} = [\theta_1, \cdots, \theta_L]^T$, over an additive white Gaussian noise (AWGN) channel. Assuming that the timing recovery is perfect without inter-symbol interference (ISI), the sampled baseband signal $\mathbf{y} = [y_1, \cdots, y_L]^T$ can be written as:

$$y_l = s_l e^{j\theta_l} + n_l = (a_l + jb_l)e^{j\theta_l} + n_l, \quad (8)$$

where $s_l$, $\theta_l$ and $n_l$ are respectively the $l$-th transmitted complex symbol ($s_l = a_l + jb_l$), the residual phase distortion and the zero mean circular Gaussian noise with variance $\sigma_n^2$.

For the data aided (DA) scenario, the transmitted symbols are independent and identically distributed (i.i.d.) and the conditional probability based on the known phase is:

$$p(\mathbf{r}|\boldsymbol{\theta}) = \prod_{l=1}^{L} p(y_k|\theta_k) = \left(\frac{1}{\pi\sigma_n^2}\right)^L \prod_{l=1}^{L} \exp\left\{-\frac{|s_l|^2 + |y_l|^2}{\sigma_n^2}\right\}\exp\left\{2\frac{\operatorname{Re}\{y_l s_l^* e^{-j\theta_l}\}}{\sigma_n^2}\right\} \quad (9)$$

For the non-data aided (NDA) case, the transmitted symbols are also i.i.d and thus $p(\mathbf{r}|\boldsymbol{\theta})$ has a similar form:

$$p(\mathbf{r}|\boldsymbol{\theta}) = \prod_{l=1}^{L} p(y_l|\theta_l)$$
$$= \left(\frac{1}{\pi\sigma_n^2}\right)^L \prod_{l=1}^{L} \sum_{s_l \in \mathbf{S}_M} \frac{1}{M} \exp\left\{-\frac{|s_l|^2 + |y_l|^2 + 2\operatorname{Re}\{y_l s_l^* e^{-j\theta_l}\}}{\sigma_n^2}\right\}. \quad (10)$$

In practice, the oscillators are never perfect and suffer from jitters. [11] and [12] have provided a mathematical model which has been widely used to describe the oscillator behavior:

$$\theta_l = \theta_{l-1} + \xi + w_l, \quad (11)$$

where $\theta_l$ is the unknown phase offset at time $l$, $\xi$ is the unknown constant frequency offset (linear drift), $w_l$ is a white Gaussian noise with zero mean and variance $\sigma_w^2$. The corresponding conditional probability can be expressed as:

$$p(\theta_l|\theta_{l-1},\xi) = \frac{1}{\sqrt{2\pi}\sigma_w}\exp\left\{-\frac{(\theta_l - \theta_{l-1} - \xi)^2}{2\sigma_w^2}\right\}. \quad (12)$$

### IV. CRBS FOR THE DYNAMICAL PHASE ESTIMATION

In practice, phase estimation can be considered using the off-line scenario and the on-line scenario. The former uses the whole observations frame and only begins estimation when the whole frame has been received, while the latter directly uses the current and previous observations. In the following, we will give both the on-line and the off-line lower bounds.

The parameters of the phase model in (9), (10) include some random parameters $\boldsymbol{\theta} = [\theta_1, \cdots, \theta_L]^T$ (i.e. the dynamical phase) and a deterministic parameter $\xi$ (i.e. the scalar linear drift). So the parameter vector can be written as:

$$\mathbf{u} = \begin{bmatrix} \mathbf{u}_r \\ \mathbf{u}_d \end{bmatrix} = \begin{bmatrix} \boldsymbol{\theta} \\ \xi \end{bmatrix} \quad (13)$$

Equation (3) thus becomes:

$$\mathbf{H}(\xi^\Delta) = E_{\boldsymbol{\theta}|\xi^\Delta}\left[\mathbf{F}(\xi^\Delta, \boldsymbol{\theta})\right]$$
$$+ E_{\boldsymbol{\theta}|\xi=\xi^\Delta}\begin{pmatrix} -\Delta_{\boldsymbol{\theta}}^{\boldsymbol{\theta}} \log p(\boldsymbol{\theta}|\xi^\Delta) & -\Delta_{\boldsymbol{\theta}}^{\xi} \log p(\boldsymbol{\theta}|\xi)_{|\xi=\xi^\Delta} \\ \left(-\Delta_{\boldsymbol{\theta}}^{\xi} \log p(\boldsymbol{\theta}|\xi)_{|\xi=\xi^\Delta}\right)^T & -\Delta_{\xi}^{\xi} \log p(\boldsymbol{\theta}|\xi)_{|\xi=\xi^\Delta} \end{pmatrix} \quad (14)$$

where $\mathbf{F}(\xi^\Delta, \boldsymbol{\theta}) = E_{\mathbf{y}|\boldsymbol{\theta},\xi=\xi^\Delta}\left[-\Delta_{\boldsymbol{\theta}}^{\boldsymbol{\theta}} \log p(\mathbf{y}|\xi,\boldsymbol{\theta})_{|\xi=\xi^\Delta}\right]$. We then decompose the HIM $\mathbf{H}$ into smaller matrices that will be useful in the sequel:

$$\mathbf{H} = \begin{bmatrix} \mathbf{H}_{11} & \mathbf{H}_{12} \\ \mathbf{H}_{21} & \mathbf{H}_{22} \end{bmatrix} = \begin{bmatrix} \mathbf{H}_{11} & \mathbf{H}_{12} \\ \mathbf{H}_{12}^T & \mathbf{H}_{22} \end{bmatrix}, \quad (15)$$

where

$$\begin{cases} \mathbf{H}_{11} = E_{\mathbf{y},\boldsymbol{\theta}|\xi=\xi^\Delta}\left[-\Delta_{\boldsymbol{\theta}}^{\boldsymbol{\theta}} \log p(\mathbf{y},\boldsymbol{\theta}|\xi)_{|\xi=\xi^\Delta}\right] + E_{\boldsymbol{\theta}|\xi=\xi^\Delta}\left[-\Delta_{\boldsymbol{\theta}}^{\boldsymbol{\theta}} \log p(\boldsymbol{\theta}|\xi)_{|\xi=\xi^\Delta}\right] \\ \mathbf{H}_{12} = E_{\mathbf{y},\boldsymbol{\theta}|\xi=\xi^\Delta}\left[-\Delta_{\boldsymbol{\theta}}^{\xi} \log p(\mathbf{y},\boldsymbol{\theta}|\xi)_{|\xi=\xi^\Delta}\right] + E_{\boldsymbol{\theta}|\xi=\xi^\Delta}\left[-\Delta_{\boldsymbol{\theta}}^{\xi} \log p(\boldsymbol{\theta}|\xi)_{|\xi=\xi^\Delta}\right] \\ \mathbf{H}_{22} = E_{\mathbf{y},\boldsymbol{\theta}|\xi=\xi^\Delta}\left[-\Delta_{\xi}^{\xi} \log p(\mathbf{y},\boldsymbol{\theta}|\xi)_{|\xi=\xi^\Delta}\right] + E_{\boldsymbol{\theta}|\xi=\xi^\Delta}\left[-\Delta_{\xi}^{\xi} \log p(\boldsymbol{\theta}|\xi)_{|\xi=\xi^\Delta}\right]. \end{cases} \quad (16)$$

#### A. Computation of $E_{\boldsymbol{\theta}|\xi^\Delta}\left[\mathbf{F}(\xi^\Delta, \boldsymbol{\theta})\right]$

From (9) (resp. (10)) in the DA (resp. NDA) scenario, $\ln p(\mathbf{y}|\xi,\boldsymbol{\theta})$ can be expanded as:

$$\ln p(\mathbf{y}|\xi,\boldsymbol{\theta}) = \ln \sum_{\mathbf{s}} p(\mathbf{y}|\xi,\boldsymbol{\theta})p(\mathbf{s}). \quad (17)$$

Using the i.i.d condition among data, the FIM can be written as:

$$E_{\boldsymbol{\theta}}\left[\mathbf{F}(\boldsymbol{\theta})\right] = J_D \mathbf{I}_L, \quad (18)$$

where $\mathbf{I}_L$ is the $L \times L$ identity matrix and $J_D$ is defined as:

$$J_D \triangleq E_{\mathbf{y},\boldsymbol{\theta}}\left[-\frac{\partial^2 \log p(y_l|\xi,\theta_l)}{\partial \theta_l^2}\right]. \quad (19)$$

Starting with the DA scenario, from (9) we have that:

$$E\left\{\frac{\partial^2 \ln p(\mathbf{y}|\boldsymbol{\theta})}{\partial \theta_l^2}\right\} = -2\cdot\text{SNR} = -\frac{2\sigma_s^2}{\sigma_n^2}. \quad (20)$$

We now turn to the NDA scenario, and from (10), one has that:

$$\frac{\partial^2 \ln p(\mathbf{y}|\boldsymbol{\theta},\xi)}{\partial \theta_l^2} = \frac{\sum_{s_l \in \mathbf{S}_M}\frac{\partial^2 p(y_l|s_l,\theta_l,\xi)}{\partial \theta_l^2}}{\sum_{s_l \in \mathbf{S}_M}p(y_l|s_l,\theta_l,\xi)} - \left(\frac{\sum_{s_l \in \mathbf{S}_M}\frac{\partial p(y_l|s_l,\theta_l,\xi)}{\partial \theta_l}}{\sum_{s_l \in \mathbf{S}_M}p(y_l|s_l,\theta_l,\xi)}\right)^2, \quad (21)$$

where $\frac{\partial p(y_l|s_l,\theta_l,\xi)}{\partial \theta_l} = p(y_l|s_l,\theta_l,\xi)\left(2\operatorname{Im}\{y_l s_l^* e^{-j\theta_l}\}/\sigma_n^2\right),$ (22)

$$\frac{\partial^2 p(y_l|s_l,\theta_l,\xi)}{\partial \theta_l^2} = p(y_l|s_l,\theta_l,\xi)\begin{pmatrix} \left(2\operatorname{Im}\{y_l s_l^* e^{-j\theta_l}\}/\sigma_n^2\right)^2 \\ -2\operatorname{Re}\{y_l s_l^* e^{-j\theta_l}\}/\sigma_n^2 \end{pmatrix}, \quad (23)$$

and $p(y_l|s_l,\theta_l,\xi)$ has been shown in (10). Unfortunately, the expectation of (21) has no simple analytical solution and one must resort to numerical methods.

#### B. Analytical Expressions of HCRBs

From (16), due to (12) and without any a priori knowledge on $\theta_l$ (*i.e.* $E_{\theta_l}\left[\Delta_{\theta_l}^{\theta_l}\log p(\theta_l)\right]=0$), we obtain matrix $\mathbf{H}_{11}$:

$$\mathbf{H}_{11}=b\begin{bmatrix} A+1 & 1 & 0 & \cdots & 0 \\ 1 & A & 1 & 0 & \vdots \\ 0 & \ddots & \ddots & \ddots & 0 \\ \vdots & 0 & 1 & A & 1 \\ 0 & \cdots & 0 & 1 & A+1 \end{bmatrix}, \quad (24)$$

where $A=-\sigma_w^2 J_D-2$ and $b=-1/\sigma_w^2$. (25)

For both the DA and NDA scenarios (see (9) and (10)), we can see that $\log p(y_l|\xi,\theta_l)$ is independent of $\xi$, *i.e.* the partial derivatives $\Delta_{\theta}^{\xi}\log p(\mathbf{y},\boldsymbol{\theta}|\xi)_{\xi=\xi^{\Delta}}$ and $\Delta_{\xi}^{\xi}\log p(\mathbf{y},\boldsymbol{\theta}|\xi)_{\xi=\xi^{\Delta}}$ are equal to 0. $\mathbf{H}_{12}$ and $\mathbf{H}_{22}$ thus become:

$$\mathbf{H}_{12}=\left[1/\sigma_w^2,\ \mathbf{0}_{1\times(L-2)},\ -1/\sigma_w^2\right]^T \quad (26)$$

$$\mathbf{H}_{22}=(L-1)/\sigma_w^2. \quad (27)$$

We now invert the HIM. Starting with $\mathbf{H}_{11}^{-1}$ and just similarly to Appendix I of [15], we find:

$$\left[\mathbf{H}_{11}^{-1}\right]_{1,l}=\frac{b^{l-1}}{|\mathbf{H}_{11}|}\left(\rho_1 r_1^{L-l-1}(r_1+b)+\rho_2 r_2^{L-l-1}(r_2+b)\right), \quad (28)$$

$$\left[\mathbf{H}_{11}^{-1}\right]_{l,l}=\frac{1}{|\mathbf{H}_{11}|}\left\{\begin{array}{l}\rho_1^2(b+r_1)^2 r_1^{L-3}+\rho_2^2(b+r_2)^2 r_2^{L-3}\\ -b^2\left(r_1^{L-2}r_2^{L-l-1}+r_1^{L-l-1}r_2^{L-2}\right)(A-2)^{-1}\end{array}\right\}, \quad (29)$$

where

$$\begin{cases} r_1=1/\sigma_w^2+\left(1-\sqrt{1+4(J_D\sigma_w^2)^{-1}}\right)J_D/2 \\ r_2=1/\sigma_w^2+\left(1+\sqrt{1+4(J_D\sigma_w^2)^{-1}}\right)J_D/2 \end{cases}, \quad (30)$$

and

$$\begin{cases} \rho_1\triangleq\left(\sqrt{1+4(J_D\sigma_w^2)^{-1}}-1-2(J_D\sigma_w^2)^{-1}\right)\Big/\left(2\sqrt{1+4(J_D\sigma_w^2)^{-1}}\right) \\ \rho_2\triangleq\left(\sqrt{1+4(J_D\sigma_w^2)^{-1}}+1+2(J_D\sigma_w^2)^{-1}\right)\Big/\left(2\sqrt{1+4(J_D\sigma_w^2)^{-1}}\right) \end{cases}. \quad (31)$$

Thanks to the block-matrix inversion formula [3], we have:

$$\mathbf{H}^{-1}=\begin{bmatrix}\mathbf{H}_{11}^{-1}+\mathbf{V}_L & -\lambda^{-1}\mathbf{H}_{11}^{-1}\mathbf{H}_{12} \\ -\lambda^{-1}\mathbf{H}_{12}^T\mathbf{H}_{11}^{-1} & \lambda^{-1}\end{bmatrix}, \quad (32)$$

where we define $\lambda\triangleq\frac{L-1}{\sigma_w^2}-\mathbf{H}_{12}^T\mathbf{H}_{11}^{-1}\mathbf{H}_{12}$ and $\mathbf{V}_L\triangleq\lambda^{-1}\mathbf{H}_{11}^{-1}\mathbf{H}_{12}\mathbf{H}_{12}^T\mathbf{H}_{11}^{-1}$.

Due to the particularly light structures of $\mathbf{H}_{11}$ and $\mathbf{H}_{12}$, we find:

$$\lambda=\frac{L-1}{\sigma_w^2}-\frac{2b^2}{|\mathbf{H}_{11}|}\left\{\rho_1 r_1^{L-2}(r_1+b)+\rho_2 r_2^{L-2}(r_2+b)+b^{L-1}\right\}. \quad (33)$$

From the definition of $\mathbf{V}_L$, the diagonal elements $\left[\mathbf{V}_L\right]_{l,l}$ can be written as:

$$\left[\mathbf{V}_L\right]_{l,l}=\left(\left[\mathbf{H}_{11}^{-1}\right]_{1,l}-\left[\mathbf{H}_{11}^{-1}\right]_{1,L+1-l}\right)^2\Big/\left(\lambda\sigma_w^4\right). \quad (34)$$

Using (29) and (34), we can then get the analytical expression of the upper diagonal elements $\left[\mathbf{H}^{-1}\right]_{l,l}$ in (32) *i.e.* the off-line HCRB associated to the estimation of $\theta_k$:

$$\left[\mathbf{H}^{-1}\right]_{l,l}=\frac{1}{|\mathbf{H}_{11}|}\left\{\begin{array}{l}\rho_1^2(b+r_1)^2 r_1^{L-3}+\rho_2^2(b+r_2)^2 r_2^{L-3}\\ -b^2\left(r_1^{L-2}r_2^{L-l-1}+r_1^{L-l-1}r_2^{L-2}\right)(A-2)^{-1}\end{array}\right\}$$
$$+\frac{b^2}{\lambda|\mathbf{H}_{11}|^2}\left\{\begin{array}{l}b^{l-1}\left(\rho_1 r_1^{L-l-1}(b+r_1)+\rho_2 r_2^{L-l-1}(b+r_2)\right)\\ +b^{L-l}\left(\rho_1 r_1^{l-2}(b+r_1)+\rho_2 r_2^{l-2}(b+r_2)\right)\end{array}\right\}^2. \quad (35)$$

Finally, just replacing $|\mathbf{H}_{11}|$ by $|\mathbf{H}_{11}(l)|$ in (35), one also obtains the analytical expression of the on-line HCRB associated to the estimation of $\theta_l$ ($l\geq 3$):

$$C_{\mathbf{H}_l}=\frac{1}{|\mathbf{H}_{11}(l)|}\left\{\begin{array}{l}\rho_1^2(b+r_1)^2 r_1^{l-3}+\rho_2^2(b+r_2)^2 r_2^{l-3}\\ -b^2\left(r_1^{l-2}r_2^{-1}+r_1^{-1}r_2^{l-2}\right)(A-2)^{-1}\end{array}\right\}$$
$$+\frac{b^2}{\lambda|\mathbf{H}_{11}(l)|^2}\left\{\begin{array}{l}b^{l-1}\left(\rho_1 r_1^{-1}(b+r_1)+\rho_2 r_2^{-1}(b+r_2)\right)\\ +\rho_1 r_1^{l-2}(b+r_1)+\rho_2 r_2^{l-2}(b+r_2)\end{array}\right\}^2, \quad (36)$$

where $\mathbf{H}_{11}(l)$ is the left-upper $l\times l$ submatrix of $\mathbf{H}_{11}$ (see (24)). We can notice that (35) and (36) do not depend on the value of the parameter $\xi$.

### C. Analytical Expressions of BCRBs

When there is no linear drift *i.e.* $\xi=0$, the parameter vector $\mathbf{u}$ contains only random parameters $\boldsymbol{\theta}$, *i.e.* $\mathbf{u}=\mathbf{u}_r=\boldsymbol{\theta}$ and the BCRB is the lower bound of the MSE. Moreover, the Bayesian information matrix (BIM) $\mathbf{B}_L$ is equal to the upper left sub-matrix of the hybrid information matrix $\mathbf{H}$:

$$\mathbf{B}_L=\mathbf{H}_{11}. \quad (37)$$

The diagonal element $\left[\mathbf{B}_L^{-1}\right]_{l,l}$ is the off-line BCRB associated to the estimation of $\theta_l$. The corresponding analytical expressions for respectively the off-line and on-line BCRB are:

$$\left[\mathbf{B}_L^{-1}\right]_{l,l}=\frac{1}{|\mathbf{B}_L|}\left\{\begin{array}{l}\rho_1^2(b+r_1)^2 r_1^{L-3}+\rho_2^2(b+r_2)^2 r_2^{L-3}\\ -b^2\left(r_1^{L-2}r_2^{L-l-1}+r_1^{L-l-1}r_2^{L-2}\right)(A-2)^{-1}\end{array}\right\}, \quad (38)$$

and $C_{\mathbf{B}_l}=\frac{1}{|\mathbf{B}_l|}\left\{\begin{array}{l}\rho_1^2(b+r_1)^2 r_1^{l-3}+\rho_2^2(b+r_2)^2 r_2^{l-3}\\ -b^2\left(r_1^{l-2}r_2^{-1}+r_1^{-1}r_2^{l-2}\right)(A-2)^{-1}\end{array}\right\}. \quad (39)$

Note that (38) (resp. (39)) is the first term on the right side of (35) (resp. (36)). The second terms in (35) and (36) represent the additional positive uncertainty brought by $\xi$ and the HCRB is always lower bounded by the BCRB.

## V. DISCUSSION

Constrained by the paper size, only the HCRB will be discussed in the following. First, one readily sees on Fig. 1 the superiority of the off-line approach compared to the on-line approach in the different positions of the block. Also, there is little improvement for the DA scenario (compared to the NDA scenario) when using a BPSK modulation but the gain becomes obvious for a larger constellation. Moreover, as the observation number increases, both the on-line and off-line CRBs decreases and tends to reach the corresponding asymptote values.

We now illustrate the behavior of the M-QAM (M=4, 16, 64 and 256) HCRB of $\theta_l$ as a function of the SNR in Fig. 2.

At high SNR (above 30dB), we notice that the various off-line CRBs logically merge and do not depend on the constellation. In this range of SNR, the received constellations are reliable enough to make correct decisions and it is sufficient to only take the present observation $y_l$ into account to estimate $\theta_l$; thus the estimation problem tends to a deterministic phase estimation problem where we estimate $L$ independent phases $\theta_l$ with $L$ independent observations.

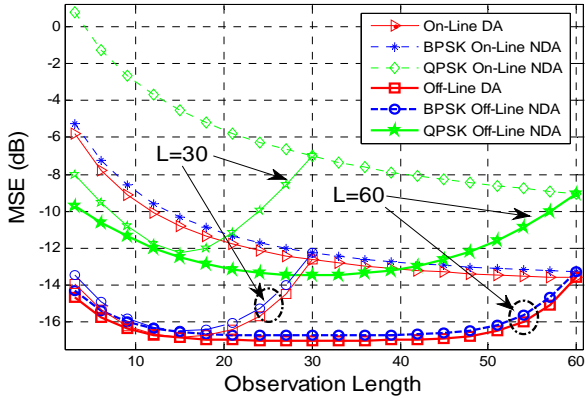

Fig. 1 HCRBs in the various block positions ($L = 30$ and $L = 60$).

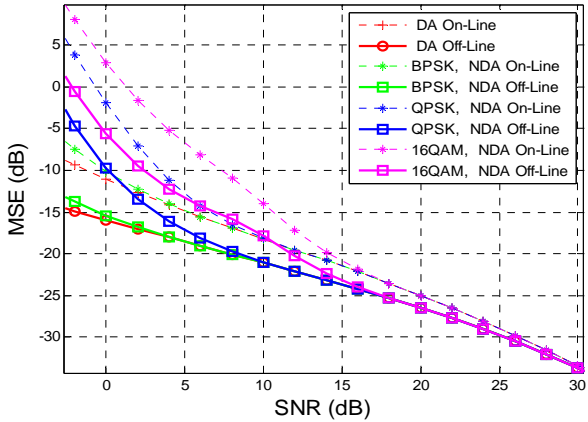

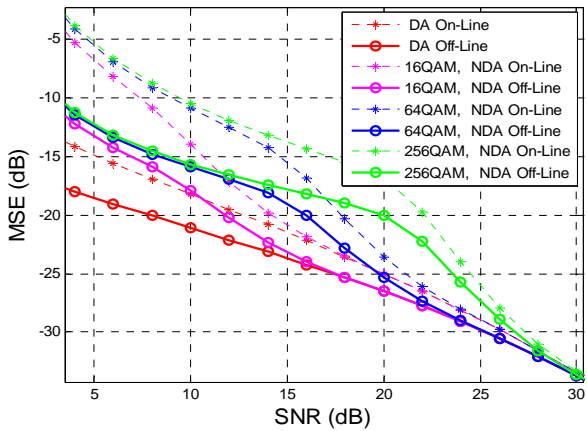

Fig. 2 HCRBs in the center of the block ($l = 30$) for various constellations.

In mid-range SNRs, one observation is not sufficient to estimate the phase offset and a block of observations can improve the estimation performance. This explains why the NDA CRBs do not merge anymore with the DA CRBs. Moreover, we notice that every time the constellation size is multiplied by 4, the threshold where the NDA bounds leave the DA bound is increased by 6dB.

At low SNR, the AWGN has more influence than the phase noise and that results in many decision errors. That is why the NDA CRBs increase quicker at low SNR than at high SNR, and particularly for the 2-dimension QAM signals (compared to the real BPSK signals).

## VI. CONCLUSION

In this paper, we have applied the general analytical form of the BCRBs and HCRBs to evaluate the performance of dynamical phase estimation. We have illustrated the phase estimation performance for arbitrary constellations. In particular, we can measure the advantage of using an off-line scenario. We also point out the difference between the QAM bounds and the traditionally studied BPSK bounds.